\documentclass[twocolumn, superscriptaddress, prb, footinbib]{revtex4-1}

\usepackage{graphicx}
\usepackage{multirow}
\usepackage{url}

\begin{document}
\title{Structural, elastic, thermal, and electronic
response of small-molecule-loaded metal organic framework materials}

\author{Pieremanuele Canepa}
\affiliation{Department of Physics, Wake Forest University,
Winston-Salem, NC 27109, USA.}
\affiliation{Department of Materials Science and
Engineering, Massachusetts Institute of Technology, Cambridge, MA 02139,
USA.}

\author{Kui Tan}
\affiliation{Department of Materials Science and Engineering, University
of Texas at Dallas, TX 75080, USA.}

\author{Yingjie Du}
\affiliation{Department of Mechanical Engineering, University of Texas
at Dallas, TX 75080, USA.}

\author{Hongbing Lu}
\affiliation{Department of Mechanical Engineering, University of Texas
at Dallas, TX 75080, USA.}

\author{Yves J. Chabal}
\affiliation{Department of Materials Science and Engineering, University
of Texas at Dallas, TX 75080, USA.}

\author{T. Thonhauser}
\email[E-mail: ]{thonhauser@wfu.edu}
\affiliation{Department of Physics, Wake Forest University,
Winston-Salem, NC 27109, USA.}

\date{\today}

\begin{abstract}
We combine infrared spectroscopy, nano-indentation measurements, and
\emph{ab initio} simulations to study the evolution of structural,
elastic, thermal, and electronic responses of the metal organic
framework MOF-74-Zn when loaded with H$_2$, CO$_2$, CH$_4$, and H$_2$O.
We find that the molecular adsorption in this MOF triggers remarkable
responses in all of these properties of the host material, with specific
signatures for each of the guest molecules.  With this comprehensive
study we are able to clarify and correlate the underlying mechanisms
regulating these responses with changes of the physical and chemical
environment.  Our findings suggest that metal organic framework
materials in general, and MOF-74-Zn in particular, can be very promising
materials for novel transducers and sensor applications, including
highly selective small-molecule detection in gas mixtures.
\end{abstract}

\pacs{68.43.Bc, 68.43.Mn, 84.60.-h, 84.60.Ve}
\maketitle

\section{Introduction}
\label{sec:intro}

External stimuli---such as heat, pressure, electric or magnetic fields,
or more complex chemical stimuli including interactions with other
chemical species---can induce changes in a material's physical
properties. The response of materials to such stimuli is the underlying
principle of sensors. The design and improvement of materials in which
this response is easy to control, reproducible, or can be related to the
interaction with specific molecules, has seen a surge of interest over
the last decade in many areas of material science.\cite{Perez02,
Saleh03, Yogeswaran08, Melde08, Tan11, Kreno12}

Many of the current commercial chemical sensors contain inorganic
semiconductor- or polymer-based films sensitive to the adsorption of
specific molecular species.\cite{Kreno12} Despite the tremendous success
of such thin-film materials, they also have shortcomings including
\emph{i)} film poisoning (which compromises the sensor lifetime),
\emph{ii)} poor selectivity toward specific molecules, \emph{iii)}
extreme operational conditions (i.e.\ high temperatures), \emph{iv)}
cross-sensitivity, \emph{v)} hysteresis (which affects the reliability
of the sensor over time), and \emph{vi)} processing and
preparation.\cite{Kreno12} In the following, we will demonstrate that
metal organic framework (MOF) materials can \emph{de facto} overcome
most of these issues with an unprecedented structural diversity,
tailorability, and tunability clearly linked to their discrete molecular
building-block nature. As a result, MOFs are studied in a large variety
of applications such as gas storage and sequestration,\cite{Nijem12a,
Canepa13} catalysis,\cite{Lee09, Luz10} polymerization
reactions,\cite{Uemura09, Vitorino09} luminescence,\cite{Allendorf09,
White09} non-linear optics,\cite{Bordiga04} magnetism,\cite{ Kurmoo09}
localized drug delivery,\cite{Horcajada09}
multiferroics,\cite{Stroppa11, Stroppa13,DiSante2013} and finally for
sensing molecules.\cite{Serre07, Soo07, Allendorf08, Tan11, Kreno12,
Qiu09, Xie10}

The flexibility of MOFs, imposed by their hybrid metal/linker type
structure, makes them very responsive to changes of the external
physical and chemical environment.\cite{Kreno12} For example, Sere
\emph{et al.}\cite{Serre07} demonstrated that IRMOF-1 displays a change
in its flexibility, about 10\%, when molecules are adsorbed within its
pores.  On the other hand, a large part of the MOF literature emphasizes
their chemical selectivity towards specific gas-molecule targets to
maximize their uptakes.  Thus, combining the flexibility of MOFs with
their selectivity towards molecules, we can build sensors with tunable
specificity.\cite{Kreno12} At the atomic level this sensing is described
as an energy transformation, i.e.\  the chemical energy released in the
adsorption process is fully (or partially) transduced into mechanical
energy, affecting the final structural, elastic, thermal, and electronic
properties of the MOF.  Therefore, in order to elucidate the link
between the change of physical properties induced by molecular
adsorption in MOFs, it is important to investigate such systems using a
bottom-up approach, i.e.\ from the atomic level to the macroscopic
level.

In this work we use DFT-based \emph{ab initio} simulations to study the
structural, elastic, thermal, and electronic response of MOF-74-Zn when
loaded with H$_2$, CO$_2$, CH$_4$, and H$_2$O. We show that---while
MOF-74-Zn is usually considered a ``rigid'' MOF compared to more
flexible candidates, e.g.\ MIL53\cite{Walker10, Ortiz12}---a substantial
change of its structural, elastic, and thermal properties is observed
after molecular adsorption.  Our predictions are supported by
\emph{in-situ} IR spectroscopy and nano-indentation measurements. We
further show that its electronic properties, and most importantly the
electron and hole effective masses, change significantly upon molecule
adsorption, with characteristic values for different adsorbates.  This
opens the door to direct sensing through electrical measurements and
leads us to conclude that slightly-doped MOF-74-Zn is a very promising
material for sensing applications.

\section{Technical details}

\subsection{Computational details}
\label{sec:sub:comp}

To model the properties presented in this work we use the van der Waals
exchange and correlation functional vdW-DF,\cite{Dion07, Thonhauser07,
Langreth09} as implemented in
VASP.~\cite{Kresse_1996:efficient_iterative,
Kresse_1999:ultrasoft_pseudopotentials} We have already successfully
applied vdW-DF to investigate the adsorption of small molecules in MOFs
and nano-structures in numerous other studies.\cite{Nijem12a, Canepa13,
Yao12, Canepa_2013:high-throughput_screening, Lopez_2013:nmr_study,
Nijem_2012:spectroscopic_characterization, Nijem_2013:water_cluster,
Tan_2012:stability_hydrolyzation, Tan_2013:mechanism_preferential,
Zuluaga_2014:study_van} Projector augmented-wave theory,\cite{Bloch94,
Kresse99} combined with a well-converged plane-wave cutoff of 600~eV,
were used to describe the wave functions. The total energy was sampled
on a 2$\times$2$\times$2 \emph{k}-point grid, resulting in four
irreducible \emph{k}-points, necessary to fully converge the stress
tensor. The density of states (DOS), band structures, and related
properties where calculated on a grid of 172 \emph{k}-points (equivalent
to a $\Gamma$-centered 10$\times$10$\times$10 \emph{k}-point mesh). The
convergence threshold for the total energy was set to 1$\times$10$^{-8}$
eV, ensuring an accurate sampling of the complex potential energy
surface of MOF-74-Zn.  The internal geometry and volume\cite{Sabatini12}
of MOF-74-Zn, empty and filled with H$_2$, CO$_2$, CH$_4$, and H$_2$O,
were fully relaxed using vdW-DF until the force criterion of
1$\times$10$^{-4}$~eV~\AA$^{-1}$ was satisfied.

\begin{figure}[t] \includegraphics[width=0.8\columnwidth]{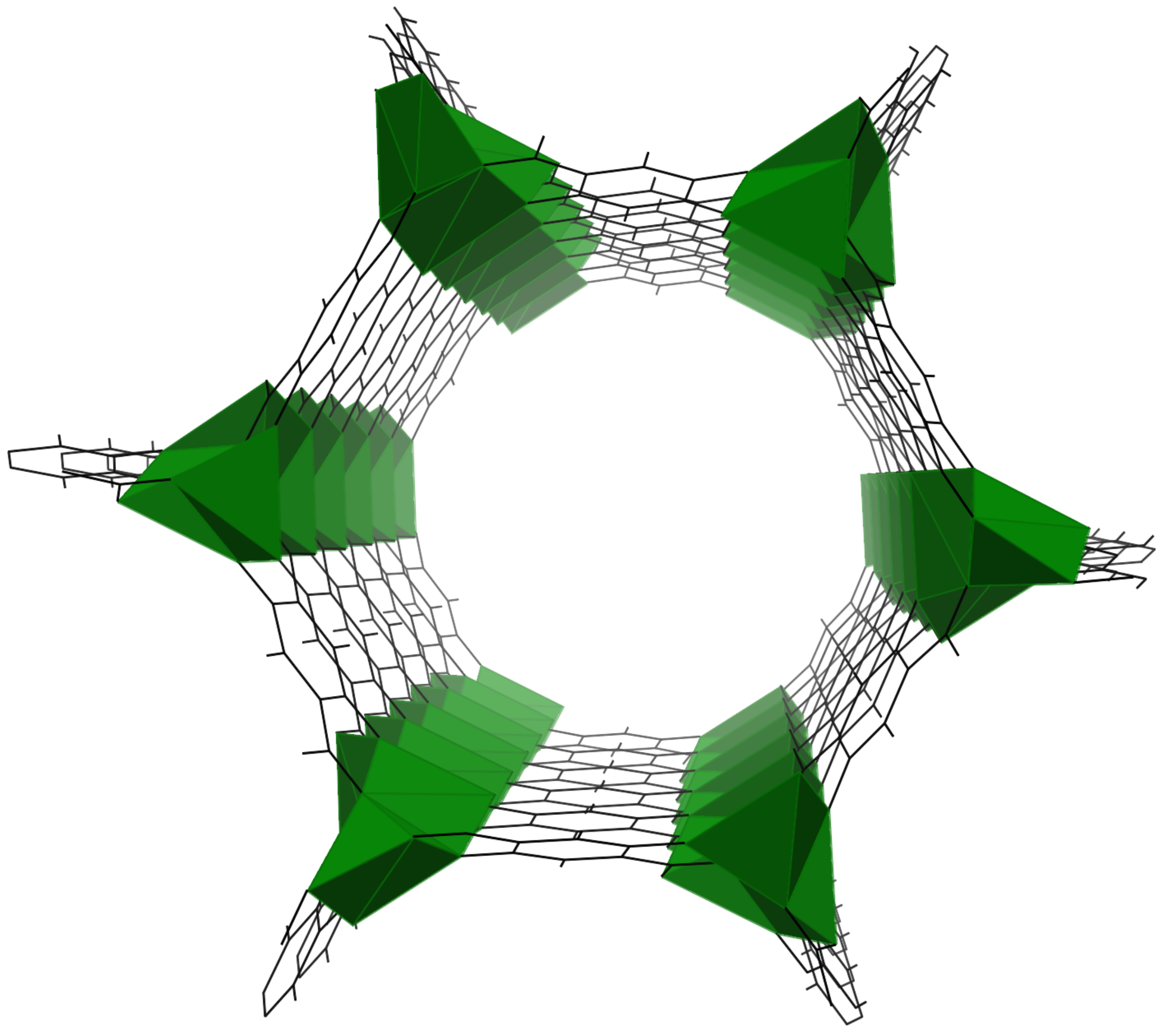}
\caption{\label{fig:structure} (Color online) MOF-74-Zn viewed along the
channel direction. Polyhedra highlight the coordination of the metal
atoms, in our case Zn atoms, which are also the primary binding sites
for small molecules inside the channel. The metal clusters at the
corners are connected by organic linkers.} \end{figure}

We start from the experimental rhombohedral structure of MOF-74-Zn with
54 atoms in its primitive cell and space group
$R\overline{3}$.\cite{Rosi05} The rhombohedral axes are
$a=b=c=15.144$~\AA\ and $\alpha=\beta=\gamma=117.778^\circ$, but the
more convenient description through hexagonal axes is $a=b=25.932$~\AA,
$c=6.836$~\AA, and $\alpha=\beta=90^\circ$ and $\gamma=120^\circ$.  For
a graphical representation of the MOF, see Fig.~\ref{fig:structure}. Six
H$_2$, CO$_2$, CH$_4$, or H$_2$O molecules are then adsorbed at the
uncoordinated zinc metal sites (six per primitive unit cell) in the MOF
nano pores, as suggested by previous X-ray and Neutron diffraction
experiments.\cite{Liu08, Dietzel08, Wu09, Dietzel06}

\subsection{Sample preparation}
\label{sec:sub:expsample}

MOF-74-Zn: A mixture of zinc nitrate hexahydrate (0.24g, 0.8mmol),
2,5-dihydroxyterephthalic (0.08 g, 0.4 mmol), 9 ml of DMF and 1 ml of
H$_2$O were transferred into a 20 ml vessel. The vessel was then  sealed
and  heated  to $120\,^{\circ}{\rm C}$  for 3 days.   After  filtering
and  washing with 20 ml of DMF,  the  product  was collected.
Successively the  product  was exchanged  with methanol  every 12 hours
during daytime  for one week to extract  the DMF solvent trapped  within
the  frameworks.  However, several previous
studies\cite{McKinlay2008,Dzubak2012} found that the  activation of the
MOF-74-Zn compound was not as easy as other MOF-74 compounds (Mg, Ni,
Co).

\subsection{Nano-indentation measurements}
\label{sec:sub:nano}

\begin{figure}[t]
\includegraphics[width=\columnwidth]{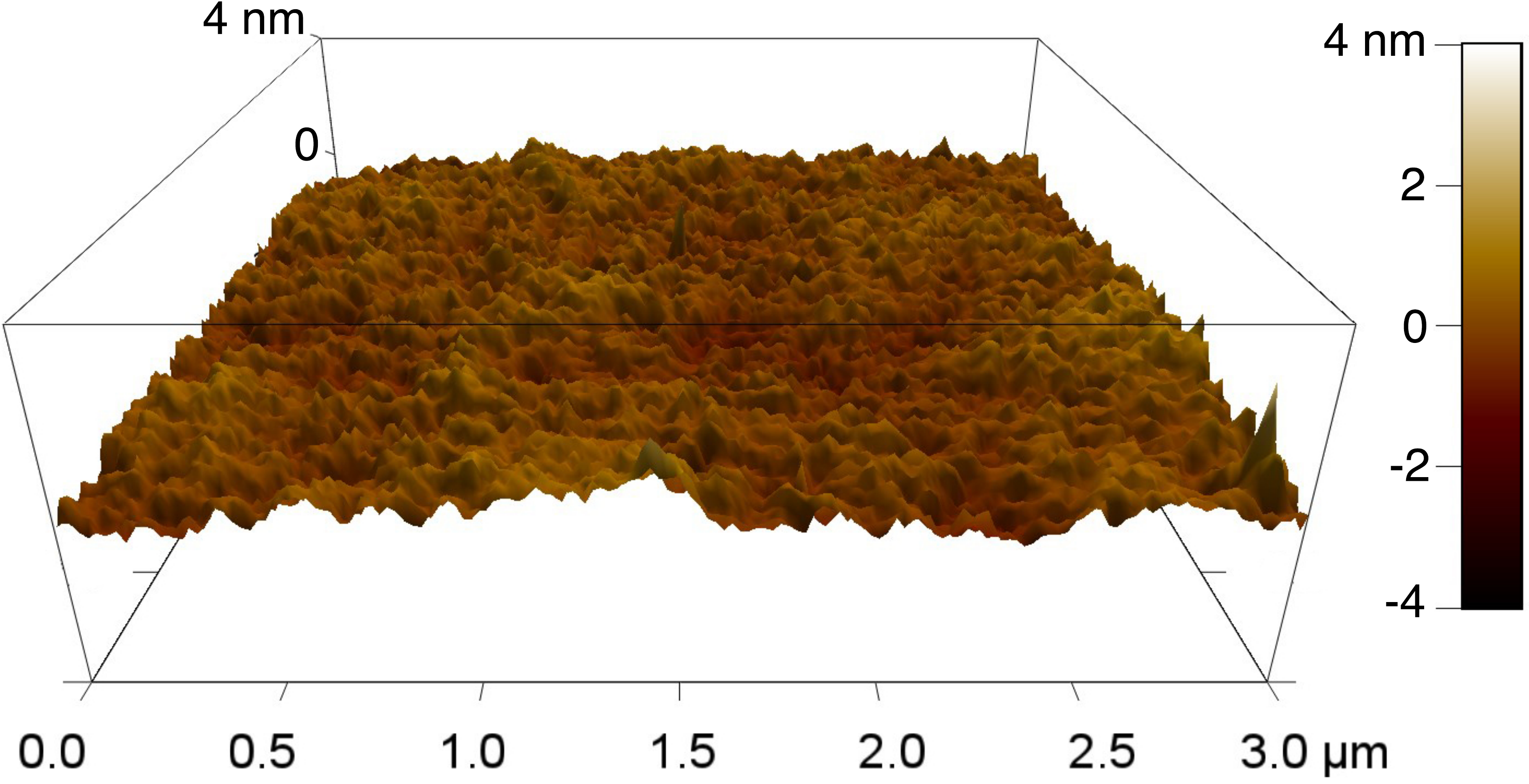}
\caption{\label{fig:microscopy} AFM topography (3 $\mu$m $\times$ 3
$\mu$m) of the MOF-74-Zn single crystal, showing a surface roughness of
less than 2 nm.}
\end{figure}

The MOF mono-crystal was larger than 20 $\mu$m in dimension and was
fixed by a 2$\mu$m epoxy thin film onto the glass slide.  Surface
roughness is critical in instrumented indentation testing. Atomic force
microscope (AFM) measurements on single crystals of MOF-74-Zn showed a
surface roughness of less than 2 nm, which is sufficient for
nano-indentation tests (see Fig.~\ref{fig:microscopy}).

An Agilent G200 Nano Indenter was used for the nano-indentation
measurements. The indenter can reach a maximum indentation depth of 500
$\mu$m (resolution: 0.2 nm) and a maximum load of 500 mN (resolution: 50
nN). A Berkovich indenter tip, made of single crystal diamond, was used
in this investigation. A maximum load of 20 mN was applied on the
indenter tip with a constant loading rate of 4 mN s$^{-1}$.
Nano-indentation tests were first conducted under ambient condition at
30\% relative humidity environment at room temperature
($24\,^{\circ}{\rm C}$). After that, the nano-indenter chamber was
sealed and dry air was purged in the chamber continuously. When the
measured relative humidity dropped below 1\%, the sample was in-situ
annealed to $150\,^{\circ}{\rm C}$ for 1 h to be fully dehydrated and
successively cooled back to room temperature ($24\,^{\circ}{\rm C}$).
Nano-indentation tests were conducted again to measure the elastic
properties of the dehydrated sample. More details about the theory
behind nano-indentation experiments are given in the Supplementary
Information.

\subsection{IR measurements}
\label{sec:sub:exp}

The IR spectra  were taken  from the  sample (2~mg) pressed on a KBr
pellet at room temperature. The pellet was placed on a high-pressure
cell purchased  from Specac at the focal point of the sample compartment
of the IR spectrometer  (Nicolet 6700, Thermo  Scientific), equipped
with a liquid N$_2$-cooled MCT-B detector. The sample was heated  at
$150\,^{\circ}{\rm C}$  in a vacuum  for 3 hours  to fully remove the
solvent  molecules.  After  cooling the  sample back to room
temperature, the spectra of the activated MOF-74-Zn were recorded under
vacuum in transmission  between  4000 and  400~cm$^{-1}$  (4 cm$^{-1}$
spectra  resolution). 760 Torr of CO$_2$ was introduced into the
pressure cell to load CO$_2$  molecules into the MOF sample, occupying
all six metal sites of the primitive unit cell.\cite{Yazaydin09}  The
high IR absorption  of gas-phase CO$_2$ leads to saturation of the
signal, therefore the cell was evacuated  and the spectra  were recorded
immediately  after evacuation  (acquisition  time = 16 sec).  The
adsorbed CO$_2$ can be easily removed in vacuum by heating the sample
slightly to $100\,^{\circ}{\rm C}$. The sample was cooled back to room
temperature ($24\,^{\circ}{\rm C}$) for hydration. 8 Torr (relative
humidity is 36\%) H$_2$O vapor was introduced to the cell to hydrate the
MOF-74-Zn sample and the spectrum was recorded after 10 min
equilibration. The H$_2$O loading is approximately 2 H$_2$O molecules
per Zn sites under this pressure.\cite{Glover11} A blank KBr pellet was
used as reference and for subtraction of gas-phase H$_2$O spectra.

\section{Results}
\label{sec:results}

\subsection{Structural response}
\label{sub:structure}

\begin{table}[t]
\caption{\label{table:lattice} vdW-DF and experimental hexagonal lattice
constants $a$ and $c$ (in \AA) and volume $V$ (in \AA$^3$) of MOF-74-Zn.
Numbers are given for the empty MOF and for the MOF loaded with six
H$_2$, CO$_2$, CH$_4$, and H$_2$O molecules. }
\begin{tabular*}{\columnwidth}{@{\extracolsep{\fill}}lcccc@{}}\hline\hline
MOF                   &                    &  $a$   & $c$   & $V$      \\\hline
\multirow{2}{*}{empty}& vdW-DF             & 26.142 & 6.875 & 4068.779 \\
                      & Exp.\cite{Rosi05}  & 25.932 & 6.836 & 3981.114 \\
\multirow{2}{*}{+H$_2$} & vdW-DF           & 26.108 & 6.816 & 4023.532 \\
                      & Exp.\cite{Liu08}   & 25.887 & 6.912  & 4011.417 \\
+CO$_2$               & vdW-DF             & 26.159 & 6.570 & 3894.077 \\
+CH$_4$               & vdW-DF             & 26.177 & 6.472 & 3840.687 \\
+H$_2$O               & vdW-DF             & 26.769 & 5.841 & 3624.784 \\\hline\hline
\end{tabular*}
\end{table}

We begin by considering the structural evolution of MOF-74-Zn when the
guest molecules, i.e.\ H$_2$, CO$_2$, CH$_4$, and H$_2$O, are adsorbed
in the structure. Table~\ref{table:lattice} compares the lattice
parameters of these models with previous experimental data.  In general,
from Table~\ref{table:lattice} we see that the vdW-DF functional tends
to overestimate both lattice parameters, i.e.\ $a$ and $c$ , affecting
the volume of the MOF, as well as the size of the nano-pores. It is well
established that vdW-DF gives slightly too large lattice parameters and
distances,\cite{Thonhauser_2006:interaction_energies} however it does
not alter the energy landscape that is most important for our analysis.
When molecules are adsorbed into the MOF structure, they produce a
change of lattice parameters and volume; the extent of this change is
solely related to the physical and chemical properties of each
adsorbate. For example, the large dipole moment of H$_2$O is responsible
for its strong interaction with the structure (see below), which
explains the significant perturbations of phonon modes induced by water
adsorption, as shown in Figure S5  in the Supporting Information.  In
the case of non-polar molecules such as H$_2$, CO$_2$, and CH$_4$ the
MOF's response depends only on the size of the molecule compared to the
MOF cavity ($\sim$13 \AA), e.g.\ CH$_4$ has a large effect on the final
lattice constants while the effect of H$_2$ is almost negligible. These
considerations are well captured by the evolution of the cavity
cross-section when several molecules are adsorbed by MOF-74-Zn.

We now move to the analysis of the adsorption energies, $\Delta E$,
which are a byproduct of the well relax geometries essential for the
calculation of the elastic and thermal properties presented in the next
sections. The adsorption energy is defined as
\begin{equation}
\label{eq:deltae}
\Delta E = E_{\rm MOF+M} - E_{\rm MOF} -E_{\rm M}{\rm(G)}\;,
\end{equation}
where $E_{\rm MOF}$ and $E_{\rm M}{\rm(G)}$ are the energies at 0 K of
the clean MOF, the molecule M in gas phase, and the adduct MOF+M, i.e.\
the product of the adsorption. Two interesting deformation contributions
$\delta _{\rm M}$ and $\delta _{\rm MOF}$---which are clearly connected
to the transduction of chemical energy at the adsorption stage into
mechanical energy---are explained in Eq.~(\ref{eq:sdelta}) and
Eq.~(\ref{eq:sdeltamof}):
\begin{eqnarray}
\delta _{\rm M}  &=& E_{\rm M,\,in\,MOF+M} - E_{\rm M}{\rm(G)}\;, \label{eq:sdelta}\\
\delta _{\rm MOF} &=& E_{\rm MOF,\,in\,MOF+M} - E_{\rm MOF}\;. \label{eq:sdeltamof}
\end{eqnarray}
Here, $E_{\rm M,\,in\,MOF+M}$ and $E_{\rm MOF,\,in\,MOF+M}$ are the
energies of the molecule and the MOF in their adsorption geometries.
$\delta _{\rm M}$ and $\delta _{\rm MOF}$ express the cost in energy
that both adsorbate and MOF have to pay during the adsorption process.
$\delta _{\rm M}$ and $\delta _{\rm MOF}$ are obtained by partitioning
the adsorption energy, and thus they are naturally enclosed in the
definition of $\Delta E$.

Table~\ref{table:binding} reports $\Delta E$, the corresponding $\Delta
E$ corrected by the zero-point energy $\Delta E_{\rm ZPE}$, the enthalpy
$\Delta H_0$ at 298 K, $\delta _{\rm M}$, and $\delta _{\rm MOF}$ for
H$_2$, CO$_2$, CH$_4$, and H$_2$O in MOF-74-Zn.\footnote[4]{The $\Delta
E_{\rm ZPE}$ correction to the adsorption energy as well as the $\Delta
H_0$ were computed on the basis of the harmonic approximation as
detailed in Ref.~\onlinecite{Sholl2009}.} Not surprising,
from Table~\ref{table:binding} we see that water binds much stronger
than the other molecules and its presence in the MOF nano-pores thus
affects the capability to effectively adsorb H$_2$, CO$_2$, or CH$_4$.
We have clarified this important aspect for the iso-structural MOF-74-Mg
in previous work.\cite{Canepa13} The distinct response of MOF-74-Zn to
water molecules was recently also demonstrated by Robinson \emph{et
al}.,\cite{Robinson12} indicating that MOFs might be ideal to detect
humidity even in small traces.

The effect of the ZPE and thermal corrections on the $\Delta E$ energy
range is only within  a few kJ~mol$^{-1}$ and does not alter the final
adsorption picture. More interesting is the effect of the deformation
contributions to the adsorption energies. $\delta _{\rm M}$ and $\delta
_{\rm MOF}$ in Table~\ref{table:binding} show that both molecules and
structures undergo a geometry reconstruction during the adsorption.  The
negative sign of $\delta _{\rm M}$ is simply due to the attractive
intermolecular interactions; their magnitudes only depend on the
molecular size and the extent of pore reconstruction, the latter being
connected to the nature of the metal ions. In general, we find that the
MOF structure is subjected to a larger structural reconstruction after
molecular adsorption---a clear indication that the MOF structure and its
elastic properties are altered.

From this preliminary analysis of the adsorption and deformation
energies we gain a qualitative picture of how the molecule-specific
adsorption process induces mechanical changes at the structural level of
the guest molecules and the MOF (see Table~\ref{table:lattice}
and~\ref{table:binding}). This has important implications for the
elastic, thermal, and electronic properties of the MOF, which, in turn,
are crucial for the development of new generation sensors.

\begin{table}[t]
\caption{\label{table:binding} $\Delta E$, $\Delta E_{\rm ZPE}$, $\Delta
H_0$ at 298.15 K, $\delta _{\rm MOF}$, and $\delta _{\rm M}$ in kJ~mol$^{-1}$
for MOF-74-Zn with six adsorbed H$_2$, CO$_2$, CH$_4$, and
H$_2$O molecules.}
\begin{tabular*}{\columnwidth}{@{\extracolsep{\fill}}lccccc@{}}\hline\hline
Model & $\Delta E$ & $\Delta E_{\rm ZPE}$ & $\Delta H_0$
& $\delta _{\rm MOF}$ & $\delta _{\rm M}$\\\hline
+H$_2$   & --20.9 & --20.3  & --19.8  & 0.9  & --1.5 \\
+CO$_2$  & --52.4 & --50.4  & --49.8  & 0.7  & --3.1 \\
+CH$_4$  & --40.1 & --38.9  & --39.3  & 0.6  & --3.3 \\
+H$_2$O  & --73.9 & --72.4  & --72.2  & 3.4  & --3.8 \\
\hline\hline
\end{tabular*}
\end{table}

\subsection{Elastic response}
\label{sec:elast}

\begin{table}[t]
\caption{\label{table:elastic} Elastic constants ($C_{ij}$), bulk ($B$),
shear ($G$), and Young's moduli ($Y$) in GPa for
empty MOF-74-Zn, as well as loaded with H$_2$, CO$_2$, CH$_4$, and
H$_2$O.  For the bulk, shear, and Young's moduli only the Hill's means
are reported. $x$, $y$, and $z$ components of the Young's modulus and
$B/G$ are also reported.}
\begin{tabular*}{\columnwidth}{@{\extracolsep{\fill}}lrrrrr@{}} \hline\hline
Prop.        & MOF &  +H$_2$  &  +CO$_2$  & +CH$_4$& +H$_2$O    \\\hline
$C_{11}$   & 14.84   &  17.00   &  19.29  & 24.16  & 31.52      \\
$C_{33}$   & 15.34   &  17.83   &  19.86  & 25.96  & 33.25      \\
$C_{44}$   & 13.03   &  12.25   &  12.68  & 12.75  & 13.47      \\
$C_{12}$   & 5.60    &   5.31   &   4.41  & 6.07   & 10.09      \\  
$C_{13}$   & 8.64    &  13.38   &  14.04  & 19.84  &  9.85      \\
$C_{14}$   & 4.89    &   2.87   &   3.66  & 1.74   &  5.17      \\
$B$        & 9.91    &  13.54   &  10.05  & 13.34  & 17.57      \\
$G$        & 4.66    &   6.48   &   7.20  & 6.90   & 12.04      \\
$B/G$      & 2.12    &   2.09   &   1.40  & 1.93   &  1.46      \\
$Y_x$      & 3.44    &   8.60   &   1.49  & 2.19   & 26.45      \\ 
$Y_y$      & 6.44    &   9.80   &   9.87  & 8.44   & 33.28      \\
$Y_z$      & 3.72    &   8.10   &   1.58  & 2.58   & 28.99      \\\hline\hline
\end{tabular*}
\end{table}

We now move to the analysis of the elastic properties of loaded
MOF-74-Zn. Table~\ref{table:elastic} reports the elastic constants and
the bulk, shear, and Young's moduli of MOF-74-Zn (see the Supplementary
Information for more details).  From this table we see that the elastic
constants and derived moduli are generally small, which is to be
expected due to the soft nature of MOFs.\cite{Tan11, Tan12, Ortiz12} We
find that the rhombohedral structure of MOF-74-Zn is stable according to
the Born\cite{Born88} stability criteria:
\begin{eqnarray}
\label{eq:stability}
C_{11} -|C_{12}| &>& 0\;,\nonumber\\
(C_{11} + C_{12})\,C_{33} &-&2C_{13}^2 > 0\;,\nonumber \\
(C_{11} + C_{12})\,C_{44} &-&2C_{14}^2 > 0\;.
\end{eqnarray}
The collective physical quantities (such as bulk, shear, and Young's
moduli) indicate the flexibility of the material in agreement with
previous literature on MOFs.\cite{Tan11,Tan12,Ortiz12} For almost all
adsorbates, the molecular adsorption in the MOF pores induces a
substantial increase of the elastic constants---i.e.\ the MOF looses
some of its flexibility---suggesting an underlying correlation between
the MOF elastic constants and its density (see below). The bulk modulus
also increases after molecular adsorption, demonstrating that the
material is less prone to compression. Remarkable are the changes in
elastic constants due to water adsorption, almost 50\%  (see
Table~\ref{table:elastic})---again demonstrating that the MOFs'
initial flexibility is strongly altered by the nature
of the adsorbing molecules. 

\begin{figure}[t]
\includegraphics[width=0.9\columnwidth]{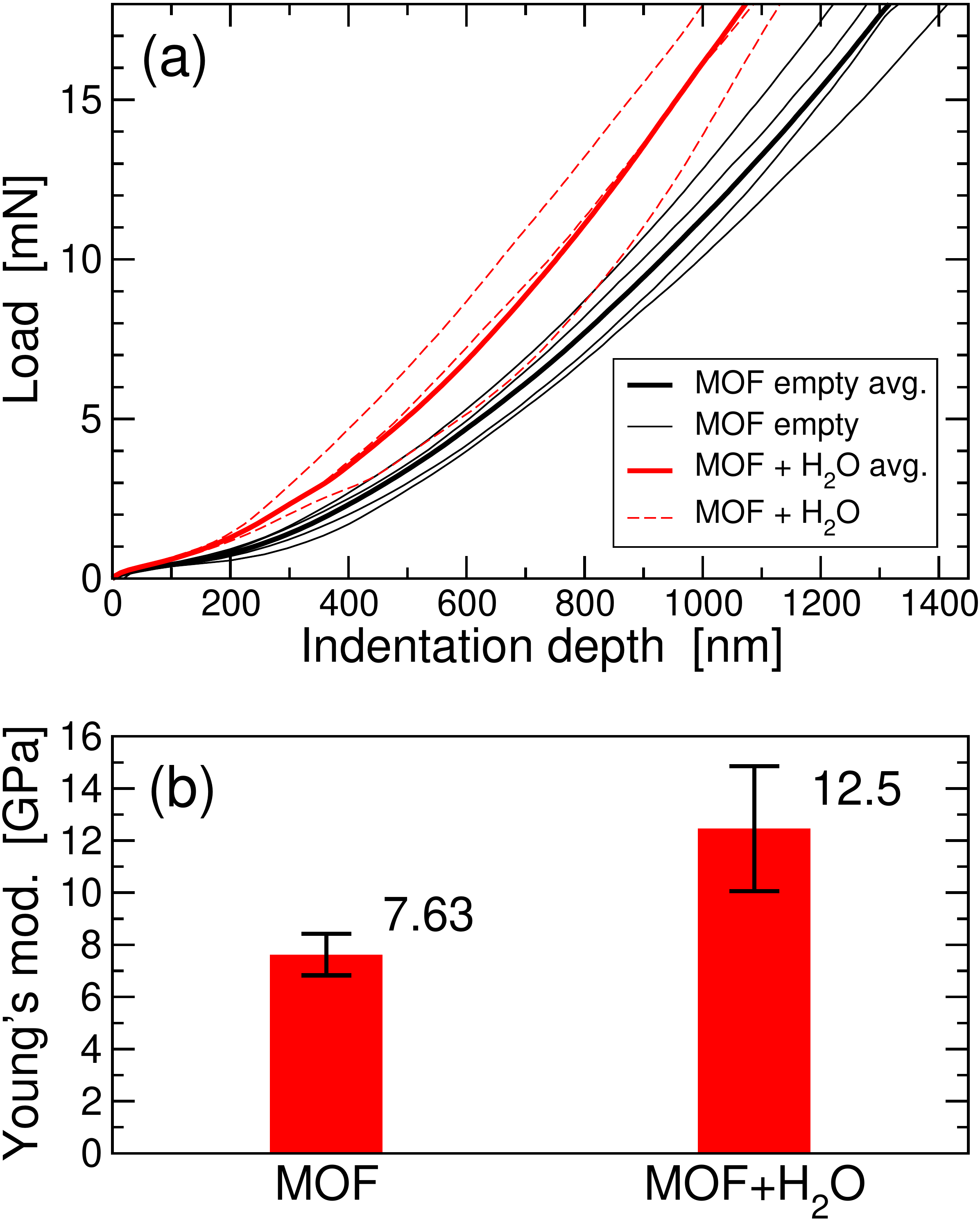}
\caption{\label{fig:expyoungmoduli} (Color online) (a) Stress-load as a
function of indentation depth of nano-indentation on the empty MOF-74-Zn
and loaded with 30\% relative humidity. Results are shown for several
repeated experiments performed on the MOF-74-Zn crystal and ``avg.''
refers to the average of those experiments.  (b) Experimental Young's
moduli of empty MOF-74-Zn and loaded with H$_2$O (30\% relative
humidity environment) and relative error bars.}
\end{figure}

An important quantity that tells us about the ductility of the MOF  is
the ratio between the bulk and the shear moduli $B/G$. Note that the
bulk modulus $B$ is generally connected to the resistance to fracture of
the material, while the shear modulus $G$ is related to the resistance
to plastic deformation; the threshold between brittle and ductile is a
value of approximatively 1.75. The values of $B/G$ in
Table~\ref{table:elastic} demonstrate that the empty MOF-74 is a rather
brittle material, but its brittleness decreases after molecular
adsorption. A measure of the stiffness of the MOF-scaffold is given by
the Young's modulus. The small values computed for the Young's moduli
are the clear indication of the flexible nature of MOF-74-Zn.  In order to
confirm our theoretically predicted trend, in-situ nano-indentation
measurements are performed on MOF-74-Zn sample in the hydrated state
(under ambient condition and relative humidity $~$30\%) and dehydrated
state, which was achieved by in-situ heating up to $150\,^{\circ}{\rm
C}$ under dry air flow.  Figure~\ref{fig:expyoungmoduli}a shows the
load-displacement data collected on the MOF-74-Zn crystal surface with a
penetration depth over 1000 nm. Note that the hydrated MOF-74-Zn sample
is measured under 30\% humidity at ambient condition.

From these measurements the elastic modulus was extracted using Oliver
Pharr's method as detailed in Sec.~\ref{sec:sub:nano} and in the
Supplementary Information. The measured averages of Young's modulus for
the empty MOF-74-Zn sample in dry air and the sample in a 30\%
relative-humidity environment are 7.63 GPa and 12.46 GPa respectively.
Although the Young's modulus of the epoxy support (5.00 GPa) can---in
theory---interfere with the elastic measurements, the MOF's
thickness-to-indentation depth ratio is more than 30, guaranteeing that
the overall effect of the epoxy substrate is negligible. In
Figure~\ref{fig:expyoungmoduli}b, the experimental nano-indentation
results (especially for dehydrated condition) are in good agreement with
calculated $Y_y$ Young's modulus components, as shown in
Table~\ref{table:elastic}. Again, the Young's moduli, similar $B$ and
$G$, are subject to change during molecular adsorption, causing an
overall stiffening of the MOF itself. Experimental and calculated
Young's moduli of MOF-74-Zn range between 6.0 -- 12.5~GPa,
characteristic of wood along the grain (11~GPa) or human cortical bone
(14~GPa).  Although our DFT calculations overestimate the Young's
modulus for the case of H$_2$O in MOF-74-Zn, the agreement between
experiment and theory is still remarkable, considering the complexity of
the system and the measurement. The difference between the experimental
data and theoretical values of the hydrated MOF can---at least in
part---be attributed to the difficulty of fully loading the sample and
thus saturating all metal sites with water
molecules.\cite{Dietzel2008,McKinlay2008,Dzubak2012}

Note that the measured results  under ambient onditions (relative
humidity 30\%) and dry air flow follow the trend of the theoretical
prediction, i.e.\ the Young's modulus increases upon water adsorption.
Unfortunately, while the elastic measurements in presence of H$_2$,
CO$_2$, and CH$_4$ have been planned, this task is not trivial,
requiring a complex setup that includes building an \emph{in-situ}
chamber around the nano-indentation head. Thus, at the moment we cannot
report measurements for those molecules.

\begin{table}[t]
\caption{\label{table:poisson} Density $\rho$ (in g cm$^{-3}$),
Poisson's ratio components $u_i(j)$, longitudinal ($v_l$), transversal
($v_t$), and mean ($v_m$) sound velocities (in Km~s$^{-1}$), and Debye
temperature (in K) for empty MOF-74-Zn, as well as loaded with H$_2$,
CO$_2$, CH$_4$, and H$_2$O. These properties were evaluated starting
from Hill's moduli. See Supplementary Information for the definition of
these quantities.}
\begin{tabular*}{\columnwidth}{@{\extracolsep{\fill}}lrrrrr@{}}\hline\hline
Prop.           & MOF        &  +H$_2$ &  +CO$_2$  &  +CH$_4$  & +H$_2$O \\\hline
$\rho$          &  1.25      & 1.21    &  1.58     &  1.77     & 1.34    \\
$u_1(2)$        &  0.45      & 0.25    &  0.33     &  0.40     & 0.27    \\
$u_1(3)$        &  0.32      & 0.93    &  0.33     &  0.31     & 0.27    \\
$u_2(3)$        &  0.11      & 0.05    &  0.12     &  0.01     & 0.14    \\
$u_m$           &  0.30      &  0.41   &  0.26     &  0.24     & 0.23    \\
$v_l$           &  3.11      &  4.28   &  3.59     &  4.28     & 5.01    \\
$v_t$           &  1.59      &  2.07   &  1.93     &  2.31     & 2.99    \\
$v_m$           &  1.78      &  2.33   &  2.16     &  2.58     & 3.31    \\
$\Theta _D$     &  170       &  215    &  188      &  226      & 296     \\
\hline\hline
\end{tabular*}
\end{table}

The stiffness of the MOF is pertinent in the uniaxial deformation of the
material defined by Poisson's ratio (Eq.~(4) in the Supplementary
Information), which we report in Table~\ref{table:poisson}.  When the
MOF structure is deformed axially, the lateral deformation is given by
the Poisson's ratio $u_i(j)$.  Usually, the lower bound of $u_i(j)$  is
$-1$, meaning that the material does not undergo lateral deformation and
maintains its original volume; while the upper bound of 0.5 corresponds
to situations where the shape of the material does not change after
deformation.  Note that Poisson's ratio is related to the type of
bonding interactions holding the material in place; for materials that
show primarily non-directional van der Waals and ionic forces, the
Poisson's ratio is on average $\sim$0.25, in agreement with the $u_i(j)$
values in Table~\ref{table:poisson}.

Also relevant is the increase in the density $\rho$, occurring when
molecules bind to the MOF's structure, indicating that the volume of the
nano-pores is decreasing (see Table~\ref{table:lattice}
and~\ref{table:poisson}). Hence, the denser the material, the larger the
bulk modulus,  in agreement with the trend of the bulk modulus observed
when the molecule is adsorbed in the MOF nano-pores.

From the above quantities we can also calculate the sound velocities
(see Table~\ref{table:poisson}). Overall, the sound velocities for
MOF-74-Zn fall in the range of other MOFs.\cite{Allendorf08,
Maltesini06, Tan12, Ortiz12, Tan11} As expected, the longitudinal sound
velocities $v_l$ are  smaller than the transversal ones $v_t$, and both
increase during the molecular adsorption---again related to the increase
in density after molecular adsorption (see Eq.~(5) and~(6) in the
Supplementary Information).

Finally, we analyze the Debye temperature $\Theta_{\rm D}$, which is
related to the rigidity of the MOF and represents the highest
temperature that can be achieved due to a single normal vibration. Note
that $\Theta_{\rm D}$ is closely connected to the thermal analysis in
Sec.~\ref{sec:thermo}. In fact, by knowing the Debye temperature one can
recompute the heat capacity at constant volume according to the Debye
model, which is in good agreement with those presented in
Fig.~\ref{fig:cv}, based on Einstein's model. The calculated
$\Theta_{\rm D}$ for these models grows whenever a molecule is adsorbed
in the MOF cages, reflecting the molecular nature of the adsorbate. 
From a qualitative point of view, when water molecules are
introduced in the MOF structure, the high-energy frequencies---i.e.\ the
water asymmetric stretch mode---triggers the increase of the Debye
temperature. The estimated $\Theta_{\rm D}$ falls at lower values for
more weakly bound molecules such as H$_2$, CO$_2$, and CH$_4$. Overall,
the rather small Debye temperatures indicate that MOF-74-Zn presents a
relatively flexible structure, even when molecules are adsorbed in it.

\subsection{Thermal response}
\label{sec:thermo}

We begin by comparing the computed heat capacities at constant volume
from theoretical and experimental IR data, $C_v$ (Eq.~(1) in the
Supplementary Information) for MOF-74-Zn---empty and with adsorbates
H$_2$, CO$_2$, CH$_4$, and H$_2$O.

\begin{figure}[t]
\includegraphics[width=\columnwidth]{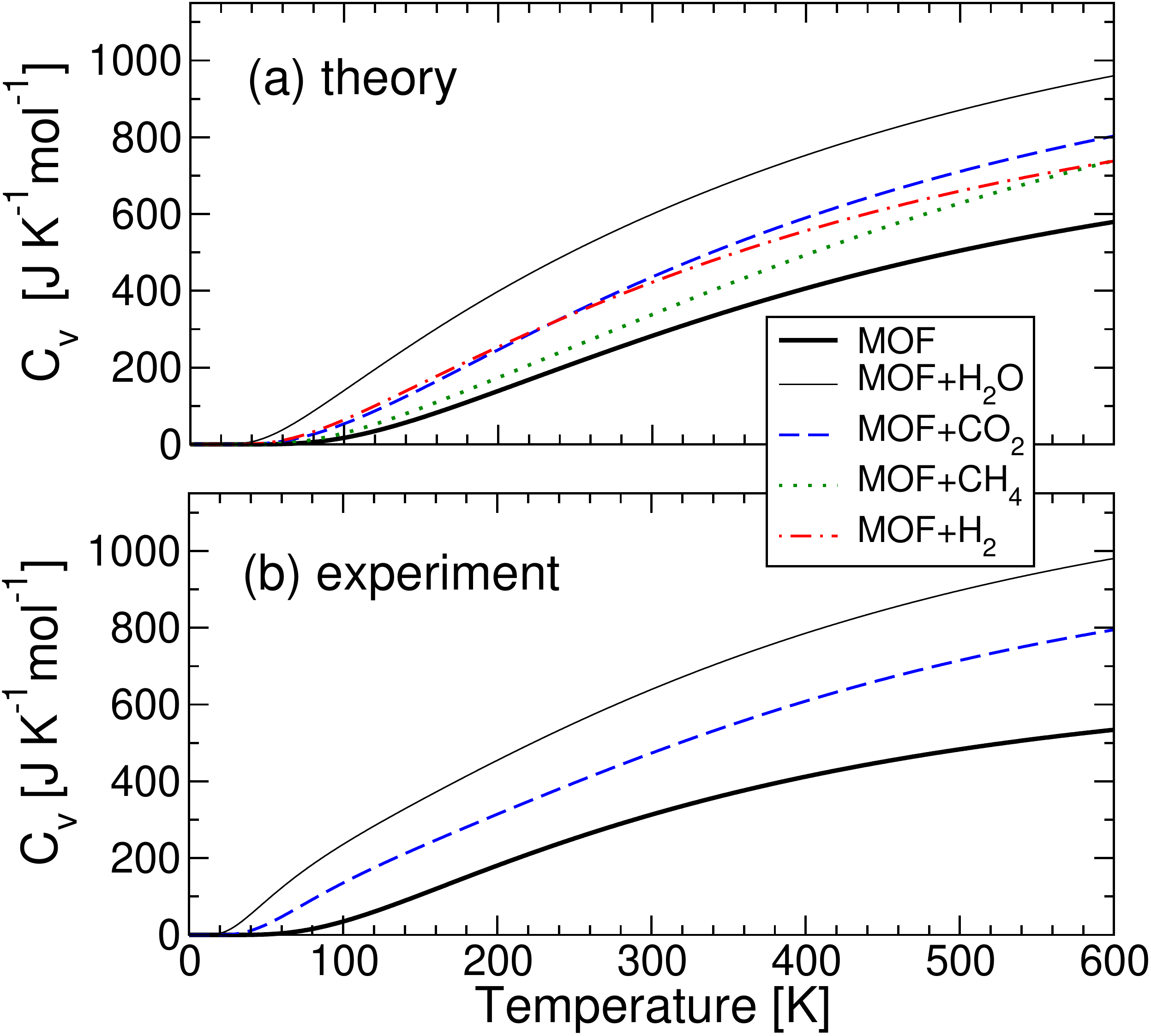}
\caption{\label{fig:cv} (Color online) (a) Constant volume heat capacity
$C_v$ calculated from our \emph{ab initio} frequencies omitting phonon
frequencies below 500 cm$^{-1}$ for the empty MOF-74-Zn, as well as
loaded with H$_2$, CO$_2$, CH$_4$, and H$_2$O. (b) same as (a), but
now calculated from experimental IR frequencies between
500$-$3800~cm$^{-1}$  (only for CO$_2$ and H$_2$O).}
\end{figure}

Figure~\ref{fig:cv}a shows the change in heat capacity computed from the
\emph{ab initio} phonon frequencies when the MOF comes in contact with
different gas molecules. Similarly, from our experimentally observed IR
frequencies of the MOF (see Figure S5 and S6 in the Supplementary
information) we can estimate the same heat capacity at constant volume,
as depicted in Fig.~\ref{fig:cv}b. Note that the experimental IR data is
only accessible in a limited spectral window (500$-$3800 cm$^{-1}$),
affecting the shape of C$_v$ in its tail at low temperatures, i.e.\ 0 --
100 K, as well as the magnitude at high temperatures, i.e.\ 600~K.  For
a better comparison between theory and experiment, we thus limit the
\emph{ab initio} frequencies included in $C_v$ to the same spectral
window, recovering excellent agreement between theory and experiment.
In general, the heat capacity increases when molecules adsorb in the MOF
nano-pores.  As with all the other responses already discussed, the
changes in the heat capacity upon loading are substantial and specific
for each adsorbate.  A clear relationship can be drawn between
adsorption energy and the trend assumed by the heat capacities---for
increasing $\Delta E$ we observe an increase of $C_v$.

\begin{figure}[t]
\includegraphics[width=1.0\columnwidth]{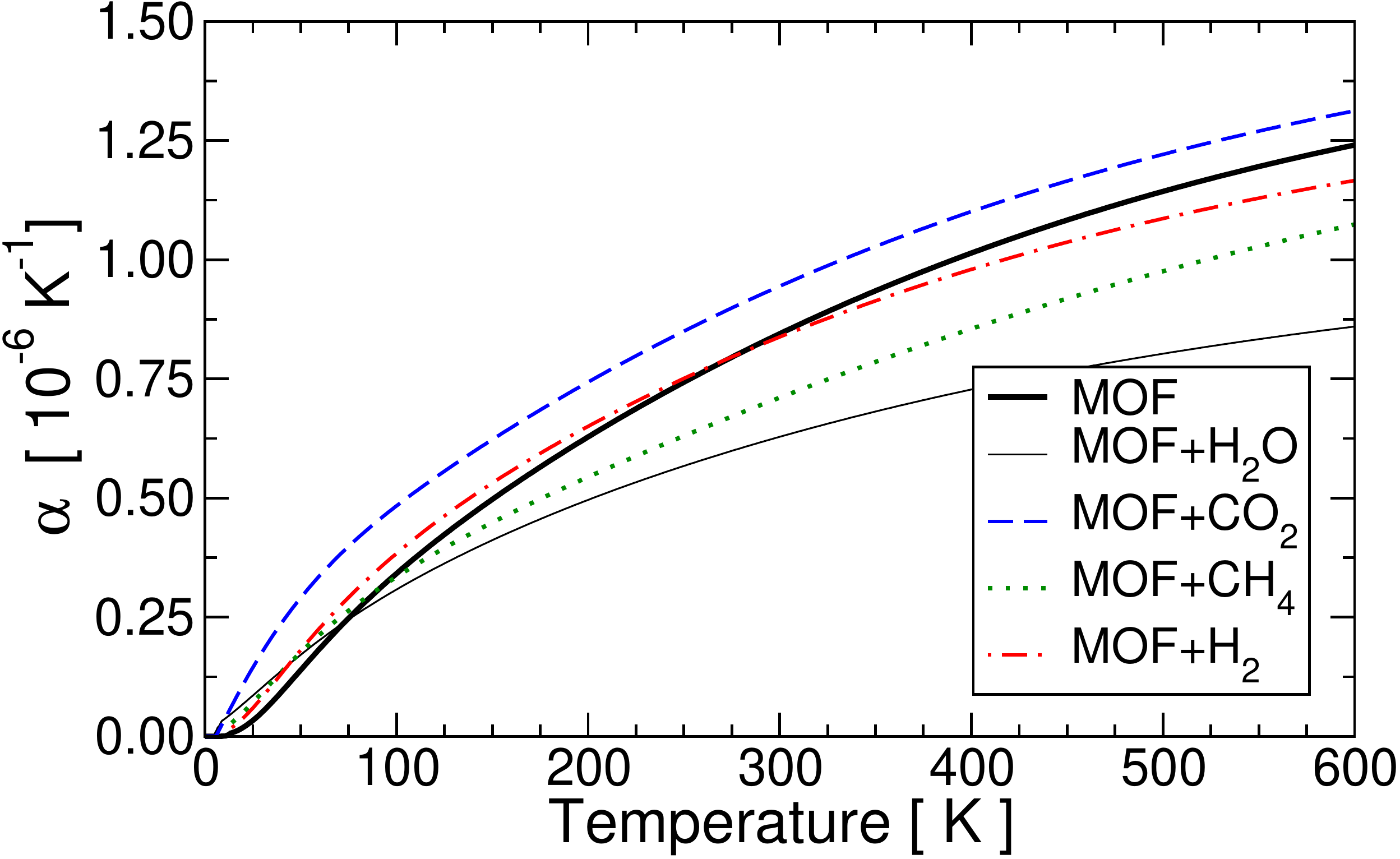}
\caption{ \label{fig:thermal} (Color online) Thermal expansion $\alpha$
as function of temperature (in 10$^{6}$ K$^{-1}$) for empty MOF-74-Zn,
as well as loaded with H$_2$, CO$_2$, CH$_4$, and H$_2$O.}
\end{figure}

$C_v$ is the main ingredient to calculate the thermal expansion $\alpha$
from Eq.~(2) in the Supplementary Information, and
Fig.~\ref{fig:thermal} plots the variation of this quantity when
molecules are adsorbed in the MOF structure.  From this figure we can
see that $\alpha$ is greatly affected by the presence of molecules in
the MOF. The extent of the thermal expansion is changed by the nature
of the guest molecule, which is related to the nature of the
interaction that occurs at the adsorption stage.

Our finding suggests that the thermal expansion of the loaded MOF-74
slowly increases with temperature compared to the empty MOF. We also
find qualitative agreement with a recent study from Queen \emph{et
al.},\cite{Queen11} who demonstrated that the adsorption of CO$_2$ in
MOF-74-Mg only induces small changes in the thermal expansion (see
Fig.~\ref{fig:thermal}). This supports that both MOF and MOF+CO$_2$ have
similar thermal expansions. Electrostatically driven adsorptions, such
as for water, introduce large changes of $\alpha$ and $C_v$, while the
adsorption of H$_2$, CO$_2$, and CH$_4$, emerging from weak van der
Waals forces, have smaller effects, though still measurable.  Note that
the magnitude of $\alpha$ also depends on the bulk modulus (see Eq.~(2)
in Supplementary Information), which is ultimately connected to the MOF
volume.  This suggests a strong correlation between $\alpha$ and the
change in volume, modulated by the magnitude of the adsorption energies
(i.e.\ the electronic characteristics of the guest molecule).  The
presence of the adsorbed molecules in the MOF cavities helps in reducing
the pore volume and thus increasing the density of the material,
resulting in a larger $C_v$ and $\alpha$.

\subsection{Electronic response}
\label{sub:electronic}

\begin{figure}[t]
\includegraphics[width=1.0\columnwidth]{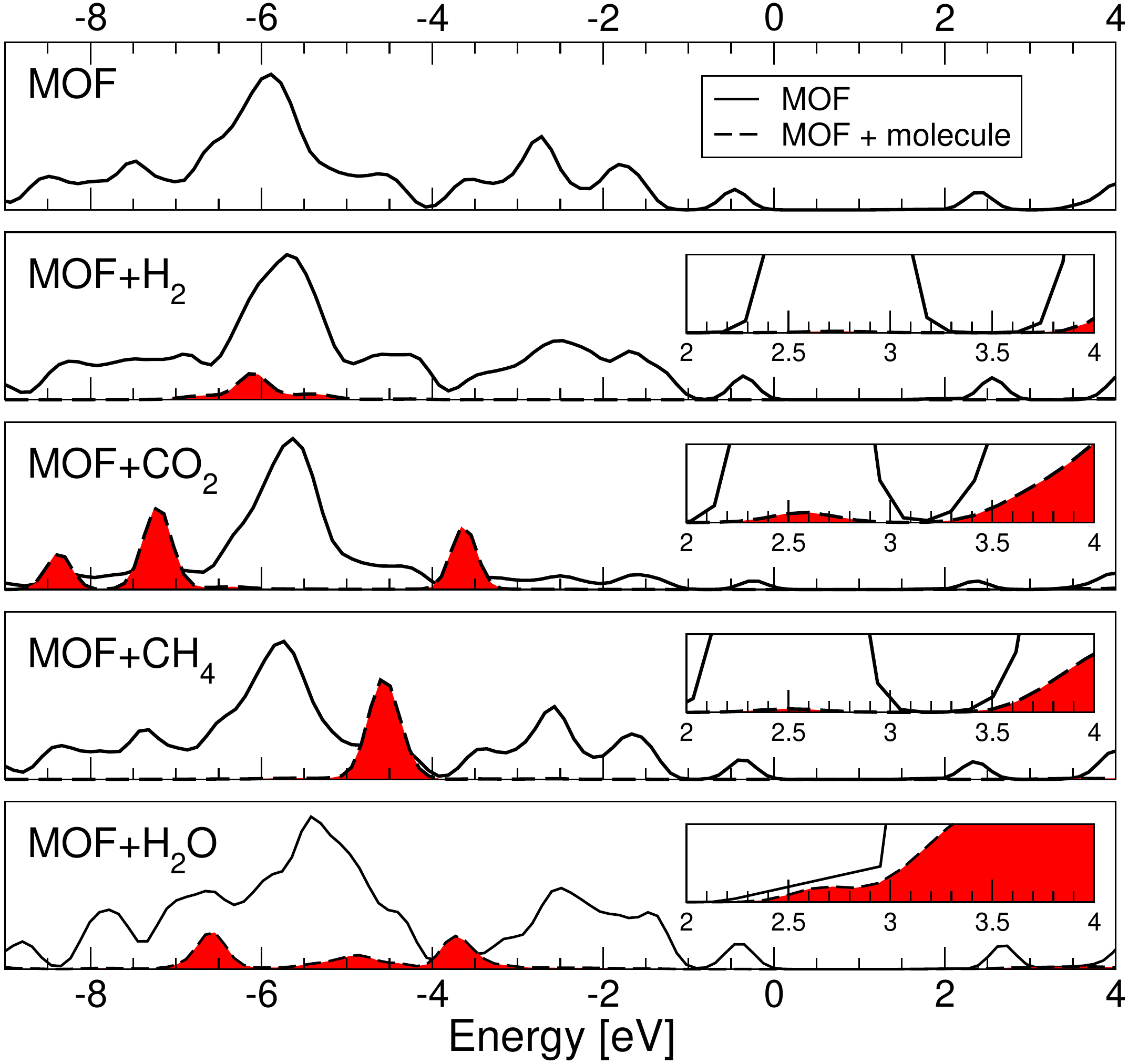}
\caption{ \label{fig:dos}(Color online) MOF and molecular projected DOS
of empty MOF-74-Zn (white), as well as MOF loaded with H$_2$, CO$_2$,
CH$_4$, and H$_2$O (red). Energies are given in eV with respect to the
top of the valence band. The insets are enlargements of the conduction
bands.}
\end{figure}

We finally come to the analysis of the electronic response of the MOF
due to molecular adsorption. Our discussion starts by considering the
alterations of the density of states (DOS) of the MOF-74-Zn after the
adsorption of the guest molecules H$_2$, CO$_2$, CH$_4$, and H$_2$O.
Figure~\ref{fig:dos} shows the MOF and adsorbate contributions to the
total DOS.

As anticipated, some molecular states (of the adsorbates) are injected
into the valence and conduction band of the MOF after the guest
molecules are adsorbed (see Fig.~\ref{fig:dos}). As such, the initial
DOS of the MOF is subjected to different changes depending on the
magnitude of the molecular interaction with the MOF scaffold; for
example, the adsorption of CO$_2$ and H$_2$O introduces changes to the
DOS of the MOF between --8~eV and --4~eV, while this is not the case of
CH$_4$ and H$_2$, which are only weakly bound to the metal sites (see
Table~\ref{table:binding}). As already suggested by the other properties
investigated above, CO$_2$ and H$_2$O introduce larger perturbations on
the MOF electronic structure.

In general, both edges of the valence and conduction bands are dominated
by the MOF states, while the molecular states lie at slightly lower and
higher energies, respectively. The small alteration of the electronic
structure introduced by H$_2$, CO$_2$, and CH$_4$ confirms that weak van
der Waals type forces hold the adsorbate to the MOF-74-Zn structure. As
shown in Fig.~\ref{fig:dos}, the molecular orbitals of H$_2$O hybridize
with the conduction bands at the edge of the band gap, slightly
affecting its magnitude.  The changes of the DOS in the conduction band for
H$_2$O (and to some extent also for CO$_2$) suggest that also the
optical properties of the MOF are strongly altered upon adsorption.

The maximum of the valence band is at the same position for the empty
MOF and for the MOF loaded with the adsorbates considered; the same
is true for the bottom of the conduction band.  In the Brillouin zone of
the rhombohedral unit cell,\footnote[5]{See
\url{http://www.cryst.ehu.es/cgi-bin/cryst/programs/nph-kv-list?gnum=148&fig=f3qra}.}
the maximum of the valence band is located at the special \emph{k}-point
$T$ and the bottom of the conduction band is along the direction $\Gamma
\rightarrow T$ and we will refer to it as $Q$.  A careful analysis of
the Brillouin zone of the MOF indicates that the vector $\Gamma
\rightarrow T$ is parallel to the MOF channel, i.e.\ the \emph{c} axis
of the corresponding hexagonal lattice.\footnote[6]{Note that the
overall symmetry (also the Brillouin zone) of the MOF remains the same
even after the six molecules are adsorbed.} 

Although DFT is not the adequate level of theory for the quantitative
prediction of electronic band gaps, our calculations suggest that
MOF-74-Zn has an indirect band gap of 2.00 eV, clearly underestimating
the experimental value of 2.83~eV  by almost 30\%.\cite{Botas11} We did
perform GW calculations to obtain a better guess for the band gap, but
those results overestimated the experimental band gap, suggesting that
exciton effects play an important role in the optical properties of
MOFs---as is not uncommon for organic materials. Solving the
Bethe--Salpeter equation is beyond the scope of this study and we thus
use our DFT results to express simple trends. The perturbation of the
electronic structure induces minor changes in the band gap magnitude
after the molecular adsorption (see Fig.~\ref{fig:dos}). Our calculated
indirect band gaps follow the trend CH$_4$ (1.99~eV) $<$ CO$_2$ (2.10
eV) $<$ H$_2$O (2.30 eV), suggesting that the stronger the adsorption,
the larger the perturbation of the MOF electronic structure with the
effect of slightly opening up the band gap. One
inexplicable exception to this trend is H$_2$, for
which we find a slightly larger band gap of 2.22~eV, nonetheless
preserving most of the electronic structure of the MOF.  Finally, from
the MOF+CH$_4$ DOS in Fig.~\ref{fig:dos} we see that the electronic
structure of the MOF itself is almost unaltered upon adsorption of
CH$_4$.

While all the above discussed responses of the MOF due to small molecule
adsorption are intriguing and useful for transducers and sensors, the
final property we discuss is probably the most relevant one. In the
remaining part we will describe how small molecule adsorption
influences the electronic conductivity, with important practical and
direct implications for sensing applications, as changes in conductivity
are easily measurable.  To this end, from the calculated band structures
above, we compute the effective masses of electrons $m^*$ and holes
$m^h$, which are inversely proportional to the corresponding
conductivities.  The effective masses are defined as the following
tensors:
\begin{equation}
\label{eq:masstensor}
m^*_{ij}\quad \text{or}\quad m^h_{ij} = \hbar^2 \left[\frac{\partial^2 E}{\partial k_i
\partial k_j}\right]^{-1}\;,
\end{equation}
where $\partial^2 E/ {\partial k_i \partial k_j}$ is the curvature in
the $i,j$ directions of the lowest and highest-lying conduction and
valence bands, respectively. These curvatures, as function of the
\emph{k}-direction, were calculated on a fine grid with a
finite-different approach (5 point sampling) employing the code by
Fonari \emph{et al.},\cite{emc} with a well converged grid-spacing of
0.025 Bohr$^{-1}$.  We have further diagonalized the effective mass
tensors, in order to get their principal-axes components. While all
calculations are performed in the rhombohedral MOF-74 unit cell, it is
more intuitive to describe the effective masses with respect to the
corresponding hexagonal representation, where the channel of the MOF is
clearly visible (see Fig.~\ref{fig:structure}). In this framework, we
can identify an effective mass $m_\parallel$ parallel to the MOF-74-Zn
channel direction, as well as two other mutually orthogonal components
$m_\perp$, which are perpendicular to the channel direction.  Results
for the effective masses in this framework are reported in Table
\ref{table:effective}.

\begin{table}[t]
\caption{\label{table:effective} Electron and hole effective masses
$m^*$ and $m^h$, parallel and perpendicular (see
Fig.~\ref{fig:structure}) to the MOF-74 channel direction (in units of
the electron mass at rest, $m_e$).  Values are reported at the bottom of
the conduction band ($Q$) and the top of the valence band ($T$) for
empty MOF-74-Zn, as well as loaded with H$_2$, CO$_2$, CH$_4$, and
H$_2$O.}
\begin{tabular*}{\columnwidth}{@{\extracolsep{\fill}}lrrr@{}}\hline\hline
$Q$      &  $m^*_\perp$ & $m^*_\perp$ & $m^*_\parallel$\\\hline 
MOF      &  214.08      & 172.30      &    2.34\\
+H$_2$   &  170.74      & 170.10      &    4.61\\
+CO$_2$  &   56.00      &   48.61     &    2.69\\
+CH$_4$  &  250.35      &  198.53     &    2.16\\
+H$_2$O  &    7.00      &    6.94     &    5.58\\\\
$T$      & $m^h_\perp$  & $m^h_\perp$ & $m^h_\parallel$ \\\hline
MOF      &    2.93      &    2.93     &    5.42\\
+H$_2$   &    3.08      &    3.08     &    9.18\\
+CO$_2$  &    3.12      &    3.10     &    5.97\\
+CH$_4$  &    2.63      &    2.63     &    5.27\\
+H$_2$O  &    2.29      &    2.29     &    4.56\\\hline\hline
\end{tabular*}
\end{table}

The components of both $m^*$ and $m^h$ in Table~\ref{table:effective}
confirm the presence of a rather large anisotropy, suggesting the
existence of more than one pathway for charge transport in the MOF
lattice, a property that is observed in many organic-based
semiconductors.\cite{Yi2012} It is interesting to see that for the
electron effective masses, the lowest value is consistently parallel to
the channel direction, while the two perpendicular components are very
similar.  On the other hand, for the hole effective masses, the lowest
value is consistently perpendicular to the channel direction, and again
the two perpendicular directions are either identical or at least very
similar.

$m^*$ and $m^h$ also vary significantly depending on the adsorbed
molecule, quantitatively supporting the picture given by the DOS of
Fig.~\ref{fig:dos}.  $m^*$ and $m^h$ also reveal that H$_2$ and CH$_4$
adsorption has a small effect on the electronic properties of the solid,
while the adsorption of H$_2$O and CO$_2$ improves overall the
electronic transport in MOF-74-Zn. Effective hole masses $m^h$ at $T$
are generally smaller than the $m^{*}$ at the bottom of the conduction
band at $Q$.  A practical guidance\cite{Fleming72} for the manufacturing
of organic semiconductors suggests that the upper useful limit for
effective masses is 25~$m_e$ (with $m_e$ the mass of the electron at
rest), suggesting that MOF-74 is of potential interest in this area.

This brings us to the question of how these important and selective
changes in conductivities can be made accessible for sensing
applications. Thermal excitement of electrons to the conduction band is
unlikely due to the large gap as well as the small
changes of the MOF electronic structure of MOF-74 upon molecular
adsorption. Optical excitations will not work either due to the
indirect band gap.  However, as we will describe below, the MOF-74
material readily lends itself to doping with other metal ions. We are
not aware of any elements that could be used for electron doping and
thus making the conduction band available, while retaining the
channel-like structure of MOF-74.  However, a whole list of transition
metal such as Cu, Ni, Mn, or Co can be used for hole doping in
MOF-74-Zn, with the advantage that MOFs built with these ions are all
isostructural and already exist.  In particular,
experiments\cite{Botas11} have confirmed that Cu and Co doping is
possible and our preliminary calculations suggest that Cu can be doped
into MOF-74-Zn at any level, until we arrive at MOF-74-Cu.  The hole
dopant will then create states in the gap just slightly above the
valence band maximum. In turn, thermal excitement creates the desired
holes at $T$, making the adsorbate-specific hole conductivities from
Table \ref{table:effective} easily accessible through electrical
measurements in novel sensor applications.

\section{Conclusions}

In summary, we elucidate the connection between the molecular adsorption
of four important molecules, i.e.\ H$_2$, CO$_2$, CH$_4$, and H$_2$O in
MOF-74-Zn and its structural, elastic, thermal, and electronic
responses.  Our vdW-DF calculations demonstrate that the chemical energy
involved in the molecular adsorption is efficiently transduced in
mechanical and thermal energy, altering the MOF properties. We further
demonstrate that the molecular adsorption induces remarkable and
adsorbate-specific changes in the MOF scaffold structure, its electronic
structure, and in its elastic and thermal properties. While MOF-74-Zn is
considered a ``rigid'' MOF, our findings attest that such a MOF can be
used for sensing a variety of important molecules with high specificity
and high molecular recognition. This study shows that by knowing the
structural, elastic, thermal, and electronic of the MOF, we can
determine the nature of the adsorbing interaction (and the molecule),
i.e.\ electrostatic or van der Waals, making MOF-74-Zn a promising
candidate for the engineering of innovative and selective sensors.


This theoretical and spectroscopic work led by TT, PC, KT, and YC was
entirely supported by the Department of Energy Grant No.\
DE-FG02-08ER46491. The nano-indentation measurements led by HL and YD
were supported by the Department of Energy Nuclear Energy University
Program (NEUP) Grant No.\ 09-416 and NSF ECCS-1307997.

\bibliography{biblio}

\end{document}